# Contact resistivity and current flow path at metal/graphene contact

K. Nagashio,[a)] T. Nishimura, K. Kita and A. Toriumi

Department of Materials Engineering, The University of Tokyo, Tokyo 113-8656, JAPAN

[a)] nagashio@ material.t.u-tokyo.ac.jp

**ABSTRACT**

The contact properties between metal and graphene were examined. The electrical measurement on a multiprobe device with different contact areas revealed that the current flow preferentially entered graphene at the edge of the contact metal. The analysis using the cross-bridge Kelvin structure (CBK) suggested that a transition from the edge conduction to area conduction occurred for a contact length shorter than the transfer length of ~1 μm. The contact resistivity for Ni was measured as ~5×10$^{-6}$ Ωcm$^2$ using the CBK. A simple calculation suggests that a contact resistivity less than 10$^{-9}$ Ωcm$^2$ is required for miniaturized graphene field effect transistors.

Graphene-based devices are promising candidates for future high-speed field effect transistors (FETs). An increase in the on/off current ratio ($I_{on}/I_{off}$) is one of the critical issues to realize the graphene FETs. Although the contact properties are important in terms of an increase in $I_{on}$, only a small number of experiments[1-6] have addressed this matter compared to the bandgap engineering for a decrease in $I_{off}$.[7,8] In fact, an ohmic contact is obtained without any difficulty due to the lack of a bandgap, but it is concerned that a very small density of states (DOS) for graphene might suppress the current injection from the metal to graphene. Recently, we reported that the contact resistivity for a typical Cr/Au electrode was high and that it varied by several orders of magnitude. It has been suggested that the contact resistivity might significantly mask the outstanding performance of the monolayer graphene channel.[5,6] Although a lower contact resistivity was reported for a Ti/Au electrode, it was described in the units of either Ωμm or Ωμm$^2$,[1,2,4] because the current flow path at the graphene/metal contact was not revealed. Furthermore, the actual contact resistivity required for FET applications has not yet been discussed. In this study, we first reveal the current flow path at the graphene/metal contact by using a multiprobe device with different contact areas. Then, the contact resistivities required for the miniaturized graphene FETs are quantitatively assessed based on the contact resistivity obtained experimentally by the cross-bridge Kelvin (CBK) method. Finally, the graphene/metal contact is discussed from the viewpoint of metal work function of contact metals employed.

Graphite thin films were mechanically exfoliated from Kish graphite onto 90 nm SiO$_2$/p$^+$-Si substrates. The number of layers was determined by the optical contrast and Raman spectroscopy.[9] Electron-beam lithography was utilized to pattern electrical contacts onto graphene. The contact metals Cr/Au (~10/20 nm), Ti/Au (~10/20 nm), and Ni (~25 nm) were thermally evaporated on the resist-patterned graphene in a chamber with a background pressure of 10$^{-5}$ Pa and were subjected to the lift-off process in warm acetone. To remove the resist residual, graphene devices were annealed in a H$_2$-Ar mixture at 300°C for 1 hour. The electrical measurements were performed in a vacuum with a source/drain bias voltage of 10 mV. The contact resistance ($R_C$) was extracted by

$$R_C = 1/2(R_{total} - R_{ch} \times L/l),$$

where $R_{total}$ is the total resistance between the source and drain, $R_{ch}$ is the channel resistance between the two voltage probes, $L$ is the length between the source and drain, and $l$ is the length between the two voltage probes, as shown in Fig. 1(b).

First, it is determined whether the contact resistivity ($\rho_C$) is characterized by the channel width ($W$) or by the contact area ($A=Wd$). Figure 1(a) shows the four-layer graphene device with six sets of 4-probe configurations (#1~#6). Ni was employed as the contact metal. The devices with different contact areas for the source and the drain were fabricated, and the contact area for the voltage probes was kept constant to avoid uncertain effects from the voltage probes, as shown in Fig. 1(b). Figure 1(c) shows the relationship between the contact area and two types of contact resistivities, $\rho_C=R_CA$ and $\rho_C=R_CW$, which were extracted by the four-probe measurements. $\rho_C$ (=$R_CA$) increases with an increasing contact area, whereas $\rho_C$ (=$R_CW$) is nearly constant for all of the devices. This indicates that $\rho_C$ is not characterized by $A$ but instead by $W$, i.e., the current flows mainly along the edge of the graphene/metal contact. In other words, the current crowding takes place at the edge of the contact metal.[10]

The current crowding should depend on the contact metal. Figure 2 shows the relationship between the contact resistivities ($\rho_C=R_CW$) and $\mu_{4P}/\mu_{2P}$ for the different contact metals Cr/Au, Ti/Au and Ni, where the two-probe mobility ($\mu_{2P}$) includes the contact resistance unlike the four-probe mobility ($\mu_{4P}$). $\rho_C$ (=$R_CW$) for Cr/Au and Ti/Au is typically high and varies largely by three orders of magnitude from ~10$^3$ to 10$^6$ Ωμm, whereas $R_CW$ for Ni is low, e.g., a minimum of ~500 Ωμm, and the variation is small. $\rho_C$ (=$R_CW$) seems to be independent of the layer number for three types of contact metals. These results suggest that the selection of the contact metal is crucially important since the outstanding performance of the graphene channel with $\mu_{4P}$ > 10,000 cm$^2$/Vs is inevitably obscured by a high $\rho_C$.

Next, to understand the edge conduction in the graphene/metal contact, the current flow path is discussed based on the transmission line model (TLM).[11] In an equivalent circuit of the TLM, there are three types of resistance: the sheet resistance of the metal ($R_M^S$), the sheet resistance of graphene ($R_{ch}^S$) and the contact resistivity ($\rho_{C\square}$). It should be noted that the unit of $\rho_{C\square}$ is defined as $\Omega\text{cm}^2$ in the TLM. The edge or area conduction can be presumed by considering the relative magnitudes of $R_M^S$ and $R_{ch}^S$. Because $R_M^S$ is smaller than $R_{ch}^S$, the current can be considered to flow preferentially in the metal to follow the least resistance path, and it enters graphene at the edge of the contact. Although a low value of $R_{ch}^S$ is expected from the high mobility of graphene, this is not the case because of the small carrier density compared to that of the metal.

In reality, however, the current does not flow just at the contact edge line. Thus, it is quite useful to estimate the effective contact distance, known as the transfer length ($d_T$). $d_T$ is approximately characterized by the relative magnitude of $R_{ch}^S$ and $\rho_{C\square}$ as

$$d_T = \sqrt{\frac{\rho_{C\square}}{R_{ch}^S}},$$

where the metal sheet resistance is neglected.[11] Hereafter, the graphene/metal contact is more accurately described by using both $\rho_{C\square}$ and $d_T$ instead of the edge-normalized $\rho_C$. To quantitatively determine $\rho_{C\square}$, the CBK structure was used.[12] The rectangular shape of monolayer graphene was prepared by using $O_2$ plasma etching. There were three electrodes on monolayer graphene. A constant current was imposed between two electrodes on the upper side, while the voltage was measured between two electrodes on the right side, as shown in Fig. 3(a). Figure 3 (b) generally shows an equivalent circuit for the CBK structure along the broken line in Fig. 3(a).[12] The circuit consists of N branches of $R_M^S$, $R_{ch}^S$, and interfacial resistance ($R_I$). It is noted that the resistances in voltage taps for the metal and channel sides are also included in this model by $R_{MT}$ and $R_{chT}$, respectively. Since no current passes out of the voltage taps, the following equation can be obtained,

$$\sum_{k=1}^{N}\frac{V_k - V_A}{R_{MT}} = 0 \text{ and } \sum_{k=1}^{N}\frac{V_k' - V_B}{R_{chT}} = 0$$

where $V_k$, $V_k'$, $V_A$ and $V_B$ are the voltages at the nodes. With the help of eq. (3), the voltage difference ($V$) between two nodes A and B is expressed as

$$V = V_B - V_A = \frac{1}{N}\sum_{k=1}^{N}(V_k' - V_k),$$

when it is assumed that the current flow is perpendicular to the interface. Thus, the total current ($I$) can be expressed as

$$I = \sum_{k=1}^{N}\frac{V_k' - V_k}{R_I} = \frac{N}{R_I}(V_B - V_A).$$

Since $R_I$ is assumed to be equivalent for all branches, $\rho_{C\square}$ is directly measured as follows,

$$\frac{1}{R_C} = \sum\frac{1}{R_I} = \frac{I}{V} = \frac{dW}{\rho_{C\square}}.$$

Figure 3(c) shows $\rho_{C\square}$ as a function of the gate voltage ($V_g$). At a high gate voltage (n=~5×10$^{12}$ cm$^{-2}$), $\rho_{C\square}$ is ~5×10$^{-6}$ $\Omega\text{cm}^2$. Furthermore, under the assumption that $R_M^S$ is much smaller than $R_{ch}^S$, the sheet resistance of graphene is required to estimate $d_T$ in eq. (2). Because the sheet resistivity of graphene was not available from the two probe geometry shown in Fig. 3(a), both the low and high mobilities that were measured previously[6,9] were used for the analysis. Figure 3(c) shows $d_T$ as a function of $V_g$ by considering high and low mobility cases. The contact length was ~4 μm, but only ~1 μm was effective for the current transfer in the present experiment. Thus, the current crowding was observed for the devices with larger contact length, as shown in Fig. 1(c). If the contact length becomes shorter than $d_T$, a transition from the edge conduction to area conduction will occur.

We next discuss the $\rho_{C\square}$ value required for miniaturized graphene FETs. Let's consider the condition on $\rho_{C\square}$ that the ratio of $R_C$ with $R_{ch}$ should be at least equal to 10 %, because the FET performance should be mainly characterized by $R_{ch}$. It can be expressed by the following equation

$$\frac{R_C}{R_{ch}} = \frac{\frac{\rho_{C\square}}{dW}}{R_{ch}^S \frac{L}{W}} = \frac{\rho_{C\square}}{R_{ch}^S dL} = 0.1.$$

Figure 4 shows $\rho_{C\square}$ required for $R_C/R_{ch} = 0.1$ as a function of $d$ for various $L$. In this calculation, a typical value for $R_{ch}^S = 250$ Ω at μ = 5000 cm$^2$/Vs and n = 5×10$^{12}$ cm$^{-2}$ was used. The dotted line which indicates the traces of $d_T$ for various $L$ was calculated by eq. (2). It separates the regions of "crowding" and "uniform injection". It is evident that $\rho_{C\square}$ becomes more severe when the contact length $d$ becomes smaller than $d_T$. For a channel length of 10 μm, the present status of $\rho_{C\square}$ for the Ni electrode satisfies the requirement. For a channel length of 100 nm, however, the required $\rho_{C\square}$ value is less than 10$^{-9}$ $\Omega\text{cm}^2$, which is four orders of magnitude lower than the present result. This value is smaller than that required for the metal/Si contact (~10$^{-8}$ $\Omega\text{cm}^2$) because the $R_{ch}^S$ of graphene is lower than that of Si. Moreover, it should be noted that $d_T$ is of the order of 10 nm.

To further decrease $\rho_{C\square}$ by four orders of magnitude, the factors that determine $\rho_{C\square}$ should be carefully considered. The work function difference (Δϕ) between graphene and metal is examined. The work functions of graphene, Ti, Cr, and Ni are 4.5, 4.3, 4.6 and 5.2 eV, respectively.[14-16] It is clear that Ni, which has the largest Δϕ, also has the lowest $\rho_{C\square}$, as shown in Fig. 2. For the case of a larger Δϕ, the electron is transferred from graphene to the metal, which will considerably increase the DOS in graphene under the metal contact and reduce $\rho_{C\square}$. Therefore, to obtain the low $\rho_{C\square}$, the metal with a larger $\Delta\phi$ is preferred.[5]

It was recently reported that the $\rho_{C\square}$ for the Ni electrode was independent of $V_g$, using the transfer length method.[3] In that method, the assumption that all of the graphene/metal contacts are equivalent must be satisfied. Although we also used this method, negative $\rho_{C\square}$ value was often extracted when $V_g$ was swept. This is because the above-mentioned method should induce a big estimation error for the case that there is a big difference between $R_C$ and $R_{sh}$. Therefore, the sharp $V_g$ dependence of $\rho_{C\square}$ determined using the CBK structure, as shown in Fig. 3(c), is more reliable because the single graphene/metal contact was measured. This $V_g$ dependence of $\rho_{C\square}$ is also explainable from the $V_g$-dependence of the DOS of graphene under the metal contact.

In summary, the contact resistance will be a limiting factor against the miniaturized graphene FETs because the $\rho_{C\square}$ should be lowered by several orders of magnitude from the present status of $\sim 5\times 10^{-6}$ $\Omega$cm$^2$. The systematic results suggest that metals with a higher $\Delta\phi$ may be preferred for achieving the low $\rho_{C\square}$ thanks to an increase in the DOS of graphene underneath the metal by the charge transfer.

**Acknowledgements**
Kish graphite used in this study was kindly provided by Dr. E. Toya of Covalent Materials Co. This work was partly supported by a Grant-in-Aid for Scientific Research from The Ministry of Education, Culture, Sports, Science and Technology, Japan.

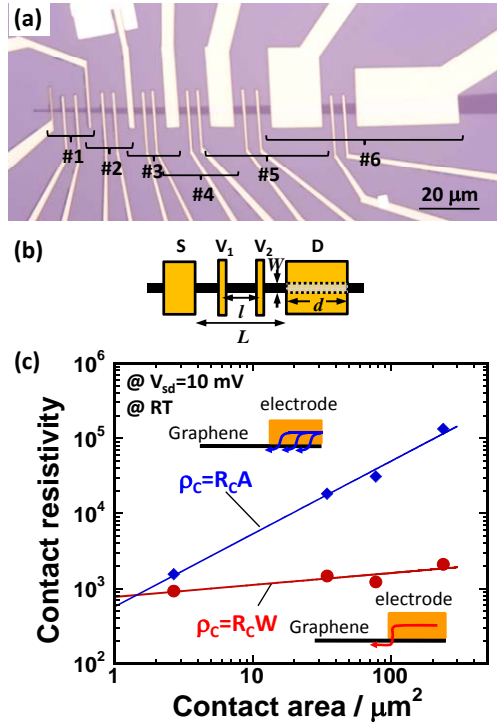

**FIG. 1** (Color online) (a) Optical micrograph of the four-layer graphene device with six sets of four-probe configurations (#1~#6). The contact metal is Ni. (b) Schematic of device. (c) Two types of contact resistivity, $R_CA$ and $R_CW$, extracted by a four-probe measurement from the devices in (a). The unit for $\rho_C=R_CA$ is $\Omega\mu m^2$, while it is $\Omega\mu m$ for $\rho_C=R_CW$.

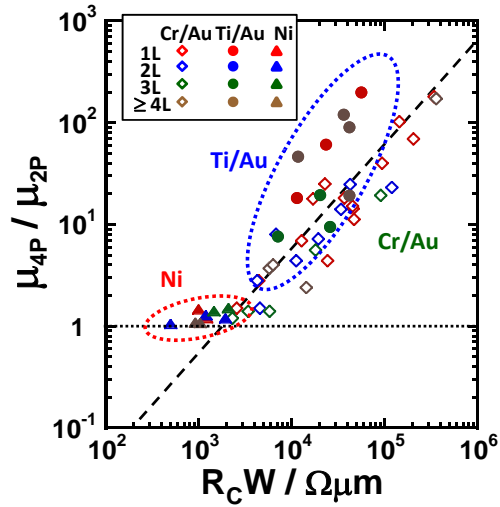

**FIG. 2** (Color online) Contact resistivities ($\rho_C=R_CW$) for the contact metals Cr/Au, Ti/Au and Ni as a function of $\mu_{4P}/\mu_{2P}$. The colors indicate the layer number.

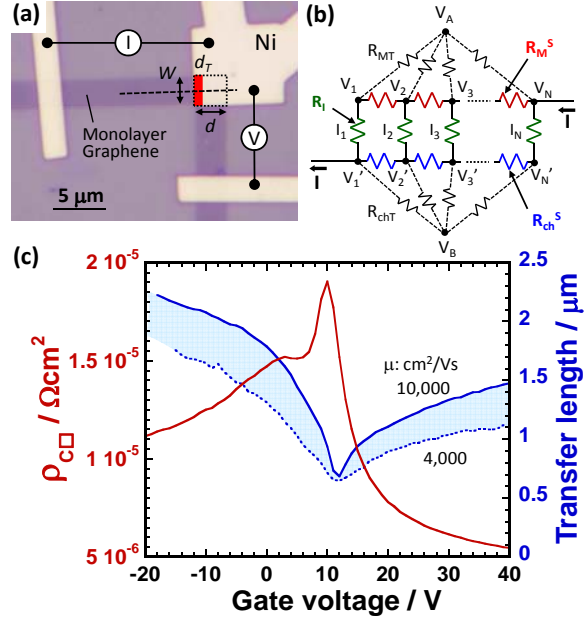

**FIG. 3** (Color online) (a) Optical micrograph of the cross-bridge Kelvin structure for monolayer graphene with a rectangular shape. (b) Equivalent circuit for the CBK structure along the dotted line in (a). The resistor symbols for $R_M^S$, $R_{ch}^S$ and $R_I$ are represented by red, blue and green colors, respectively. $V_N$, $V_N'$, $V_A$ and $V_B$ are the voltages at each node, while $I_N$ is the current between two nodes for each branch, respectively. (c) $\rho_{C\square}$ and $d_T$ as a function of the gate voltage.

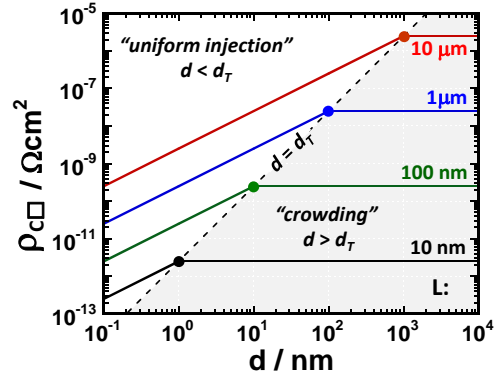

**FIG. 4** (Color online) $\rho_{C\square}$ required for $R_C/R_{ch} = 0.1$ as a function of $d$ for various $L$. In this calculation, a typical value for $R_{ch}^S = 250\ \Omega$ at $\mu = 5000\ \mathrm{cm^2/Vs}$ and $n = 5\times10^{12}\ \mathrm{cm^{-2}}$ was used. $\rho_{C\square}$ is constant for $d > d_T$ (crowding), while it decreases for $d < d_T$ (uniform injection).

5